\documentclass[%
 reprint,
superscriptaddress,
 amsmath,amssymb,
 aps,
 prl,
floatfix,
]{revtex4-2}

\usepackage{graphicx}
\usepackage{dcolumn}
\usepackage{bm}
\usepackage{subfigure}
\usepackage{geometry}
\usepackage{amsmath}
\begin{document}

\preprint{APS/123-QED}

\title{Shear viscosity scaling of granular suspensions across dilute to dense regimes}

\author{Zaohui Zhang}
\affiliation{College of Environmental and Resource Sciences, Zhejiang University, 866 Yuhangtang Rd, Hangzhou 310058, China\\}
\affiliation{Key Laboratory of Coastal Environment and Resources of Zhejiang Province (KLaCER), School of Engineering, Westlake University, 600 Dunyu Rd, Hangzhou, Zhejiang 310030, China\\}

\author{Teng Man}
\email{manteng@westlake.edu.cn}
\affiliation{Key Laboratory of Coastal Environment and Resources of Zhejiang Province (KLaCER), School of Engineering, Westlake University, 600 Dunyu Rd, Hangzhou, Zhejiang 310030, China\\}

\author{Herbert E. Huppert}
\affiliation{Institute of Theoretical Geophysics, King's College, University of Cambridge, King's Parade, Cambridge CB2 1ST, United Kingdom\\}

\author{Sergio Andres Galindo-Torres}
\email{s.torres@westlake.edu.cn}
\affiliation{Key Laboratory of Coastal Environment and Resources of Zhejiang Province (KLaCER), School of Engineering, Westlake University, 600 Dunyu Rd, Hangzhou, Zhejiang 310030, China\\}

\date{\today}
\begin{abstract}
In this letter, following an extensive experimental validation, we perform constant-volume shearing simulations of non-Brownian granular suspensions using the discrete element method coupled with the lattice Boltzmann method. We choose a wide range of solid fractions, shear rates, fluid viscosities, particle sizes, and inter-particle frictional coefficients to obtain a scaling solution for the viscous behavior of suspensions in both dilute and dense regimes. This letter demonstrates that, with a proposed dilute-dense transitional solid fraction, $\phi_d$, there exists a strong correlation between the inverse relative viscosity and the shear stress. This work incorporates both the $\phi$-dependence and the $\dot{\gamma}$-dependence of suspension viscosity in a universal framework, which provides a scaling solution for granular suspensions across dilute and dense regimes and sheds light on the dilute-dense transition mechanisms. 
\end{abstract}

\maketitle

Granular suspensions are significant for industrial production and environmental protection, but establishing a universal model for them remains challenging \cite{guazzelli2018rheology}. As a complex system combining liquid and solid, granular suspensions experience transitions from Newtonian to non-Newtonian across dilute to dense regimes, and ultimately lead to a jamming transition at a critical solid fraction $\phi_c$ \cite{liu2010jamming,guazzelli2018rheology}. Without temperature effects, the viscosity of non-Brownian suspensions with rigid spheres depends only on the solid fraction $\phi$ \cite{peters2016direct}. In a traditional viscosity scaling, the relative viscosity $\eta_r$ increases with $\phi$ and diverges at $\phi_c$, which varies from case to case \cite{guazzelli2018rheology}, providing a macroscopic description of suspension behaviors \cite{einstein1911berichtigung,batchelor1970stress,stickel2005fluid}. Based on the $\mu_e-I$ rheology of the dry granular assembly, the dimensionless viscous number $I_v$ was applied for suspensions, establishing a connection between the macroscopic behavior and the development of microscopic constitutive relationships of suspensions \cite{boyer2011unifying}. Following these, the $\mu_e-K$ rheology was proposed with a new dimensionless number $K=I_v+\alpha I$, describing the transition from a viscous to an inertial regime \cite{trulsson2012transition,tapia2022viscous}. These models work well and provide insights into the physical mechanisms of the stress state. But more investigations are needed to obtain a viscosity scaling of granular suspensions across dilute, dense, and jamming regimes.

The rheological behavior of granular suspensions varies with respect to the distance between $\phi$ and $\phi_c$ \cite{o2003jamming,hatano2010critical}. Dilute suspensions, whose solid fraction is much smaller than $\phi_c$, are often Newtonian-like, whereas concentrated suspensions with $\phi$ close to $\phi_c$ exhibit strong non-Newtonian behaviors. For dry granular assemblies, the shear stress $\tau$ is a function of the shear rate $\dot{\gamma}$ after being rescaled by ${|\phi_c-\phi|}$, which is characterized as $\tau/{|\phi_c-\phi|}^{\beta} \sim \dot{\gamma}/{|\phi_c-\phi|}^{\Delta}$. $\beta$ and $\Delta$ are constant exponents, which range from 1 to 3 \cite{dinkgreve2015universal,paredes2013rheology}. In this case, the rheological properties are classified into two regimes, an unjammed and a jammed regime, with two master curves separated by $\phi_c$. Compared with the scaling of shear stress, few studies have worked on the scaling of the shear viscosity with $|\phi_c-\phi|$ of granular materials \cite{olsson2007critical,otsuki2009universal}. \citet{olsson2007critical} simulated dry disks with different stress conditions and found a similar relationship between the inverse relative viscosity, ${\eta_r}^{-1}$, and the shear stress, $\tau$, after both being rescaled by $|\phi_c-\phi|$. However, the research on granular suspensions remains as an open question. Therefore, we conduct numerical simulations of non-Brownian granular suspensions with frictional spheres, aiming to establish a universal model across dilute to dense regimes based on critical scaling of the shear viscosity and the shear stress. 

We conduct numerical simulations of granular suspensions using a coupled discrete element method (DEM) and lattice Boltzmann method (LBM) \cite{galindo2013coupled}. The simulation code was validated with rheometer experiments as explained in the supplementary material. In the discrete element method, the solid particles are distinct elements, and their interactions are modeled by a standard linear dashpot approach. Particles contact with a limited overlapping distance $\delta_n=R_i+R_j-R_{ij}$, where $R_i = R_j = R$, $R$ is the average radius of the particles, and $R_{ij}$ is the distance between the centers of two particles \cite{cundall1979discrete}. For the fluid phase, we employ the D3Q15 model \cite{hecht2010implementation} in the lattice Boltzmann method. The simulation domain is divided into small cells, where the velocity and force are calculated based on the distribution function along 15 directions in each cell. The solid-fluid interaction in the coupled method is accounted for with the momentum change due to the collision of the DEM spheres and the LBM cells. Additionally, periodic boundary conditions are applied in the x and y directions. 

\begin{figure}
    \centering
    \includegraphics[scale=0.3]{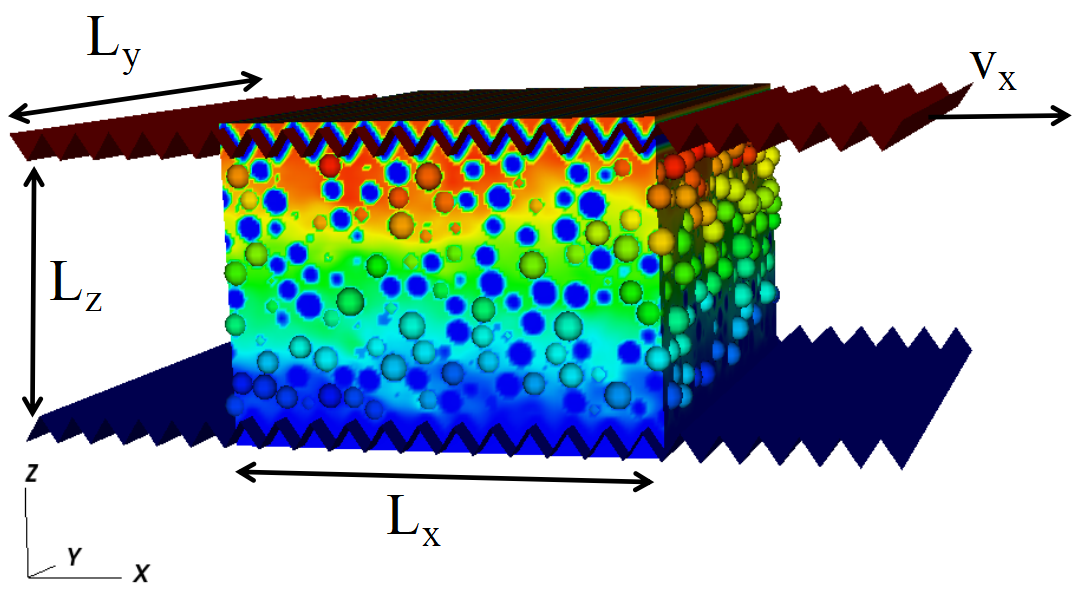}
    \caption{The DEM-LBM simulation model of parallel plate shearing, where different colors indicates the velocity profile, which increases along the height from blue to red.}
    \label{fig: DEM-LBM simulation model}
\end{figure}

We consider granular suspensions with various solid fractions $\phi$, shear rates $\dot{\gamma}$, fluid viscosities $\eta_f$, particle frictional coefficients $\mu_p$, and mean particle sizes $D_p$. We focus on suspensions with monodispersed particles that exhibit neutral buoyancy, as the solid particles and the fluid solvent have the same density ($\rho_p = \rho_f = 1.165$ g/cm$^3$). In the simulation, the average particle size $D_p$ is controlled by the maximum value $D_{max}$ and the differential coefficient $\alpha = 0.8$, while the particle size distributes in the range from $D_{max}\times\alpha$ to $D_{max}$. As shown in Fig \ref{fig: DEM-LBM simulation model}, the suspension volume remains constant $V = L_x \times L_y \times L_z$. The top plate moves in the x-direction with a given velocity $v_x$, and the shear rate $\dot{\gamma} = v_{x} / L_z$, where $L_z$ is the height of the top plate. Typically, the size ratio between the simulation space and the particle size is the same for suspensions with different particle sizes, so the number of particles $N_{ps}$ in the suspension only changes with the solid fraction. The simulation parameters, including the domain size ($L_x, L_y, L_z$), the cell size $d_x$, the stiffness values $k_n$ and $k_s$, and the time step $d_t$, are adjusted according to the average particle size $D_p$.

\begin{figure}[ht!]
\subfigure[]{\includegraphics[scale=0.28]{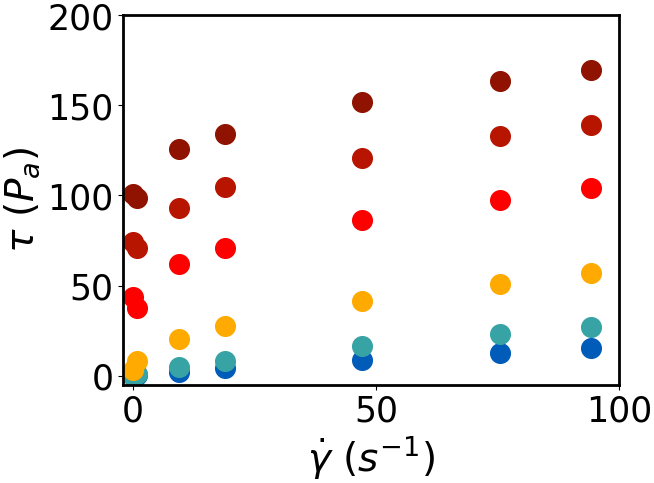}}
\subfigure[]{\includegraphics[scale=0.225]{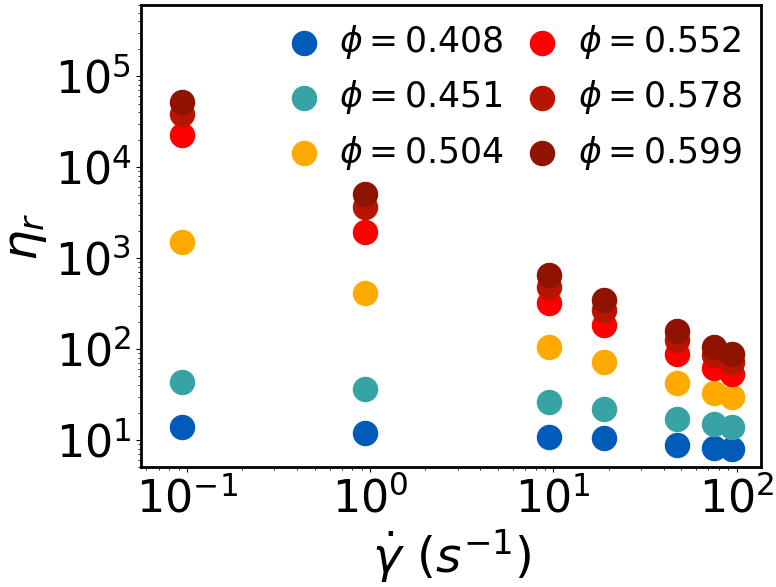}}
\subfigure[]{\includegraphics[scale=0.23]{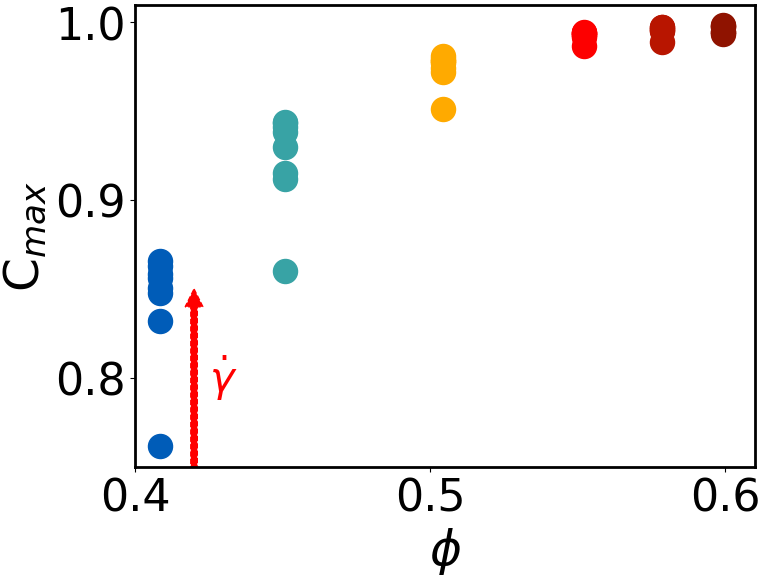}}
\subfigure[]{\includegraphics[scale=0.23]{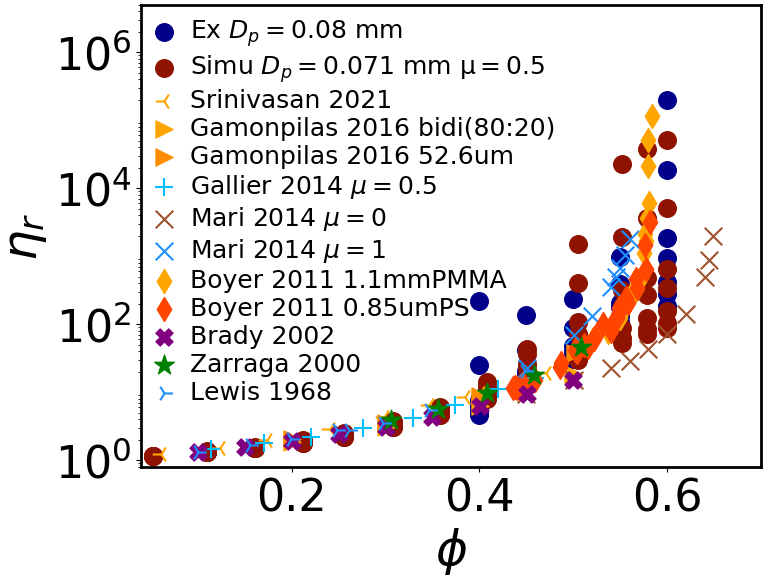}}
\caption{\label{fig: Simulation results_eta1mu05d008}The results of granular suspensions with $D_p$ = 0.071 mm, $\eta_f$ = 0.0206 $\rm{Pa \cdot s}$, and $\mu_p$ = 0.5. (a) The shear stress $\tau$ versus the shear rate $\dot{\gamma}$. (b) The relative viscosity $\eta_r$ versus $\dot{\gamma}$. (c) The maximum cluster ratio $C_{max}$ versus the solid fraction $\phi$, where the arrow indicates the growth of $\dot{\gamma}$. (d) The relative viscosity $\eta_r$ as the function of the solid fraction $\phi$, where simulation results (red dots) are compared with experiments (blue dots) and data from previous papers \cite{srinivasan2021numerical,gamonpilas2016shear,gallier2014rheology,mari2014shear,boyer2011unifying,sierou2002rheology,zarraga2000characterization,lewis1968viscosity}. The black dash-dot line represents the trajectory of $\eta_r$ in previous papers, while the blue dashed line and the red dashed line characterize trajectories of experimental results and simulation results with a finite $\phi$.}
\end{figure}

Fig \ref{fig: Simulation results_eta1mu05d008} shows the results for suspensions with $D_p = 0.071 \ \rm{mm}$, $\eta_f = 0.0206 \ \rm{Pa \cdot s}$, and $\mu_p = 0.5$, with different solid fractions $\phi$ and shear rates $\dot{\gamma}$. Generally, the rheological properties of suspensions exhibit a strong $\phi$-dependence, with the shear stress $\tau$ and the relative viscosity $\eta_r$ being proportional to the solid fraction (Fig \ref{fig: Simulation results_eta1mu05d008}(a) and \ref{fig: Simulation results_eta1mu05d008}(b)). Notably, there is a significant growth of $\tau$ and $\eta_r$ at $\phi = 0.504$ and $\phi = 0.552$. At the same time, suspensions display a pronounced $\dot{\gamma}$-dependence. The continuous shear thinning is observed in the simulation, with $\eta_r$ decreasing with increasing $\dot{\gamma}$. When $\phi > 0.5$, suspensions show non-Newtonian behavior, when $\phi \geq 0.552$, suspensions behave as Herschel-Bulkley fluids with dramatic yield stress. This behavior transition is associated with the development of the particle cluster within the suspension, which is shown in Fig \ref{fig: Simulation results_eta1mu05d008}(c). Here, $C_{max} = N_{pc} / N_{ps}$ represents the maximum size ratio of the particle cluster, where $N_{pc}$ is the number of particles in the cluster, and $N_{ps}$ is the total number of particles in the suspension. Under the shearing effect, solid particles collide with each other, resulting in the formation of particle clusters with contact networks. As the solid fraction $\phi$ increases, the total number of particles $N_{ps}$ in the suspension also increases. Within the limited space, a larger proportion of particles become involved in clusters, leading to more frequent collisions and enduring frictional contacts, and ultimately resulting in higher stress and viscosity. With finite $\phi$, both the shear stress $\tau$ and the relative viscosity $\eta_r$ change with the shear rate $\dot{\gamma}$. For $\phi < 0.5$, $C_{max}$ increases with $\dot{\gamma}$ as more particles become part of the cluster under stronger shearing effects, and the shear stress seems to increase linearly with $\dot{\gamma}$. At $\phi = 0.504$, more than 95$\%$ of the particles are involved in the cluster, reaching a densely packed state, while the $C_{max}$ only increases slightly with shear rate. In this case, the growth of particle contacts within the cluster decreases, resulting in a slight increment in stress with the shear rate, which corresponds to the shear-thinning behavior of suspensions. As $\phi$ keeps rising, particles generate a relatively stable structure by the contact network, and the yield stress $\tau_y$ appears at the onset of shearing.

The shear thinning phenomenon is more evident when we examine the development of the relative viscosity $\eta_r$ with respect to the shear rate $\dot{\gamma}$. For $\phi < 0.5$, $\eta_r$ is insensitive at low $\dot{\gamma}$, and only slightly decreases at high $\dot{\gamma}$. This indicates that even before the dense packing state, strong shearing can induce non-Newtonian behavior of suspensions. Upon reaching the dense packing state, the initial relative viscosity becomes large due to the particle contact structure, and $\eta_r$ significantly decreases as the structure is disrupted by shearing. With $\dot{\gamma}$ increasing, $\eta_r$ continues to decrease, but with small increments in stress. We investigate the development of $\eta_r$ as a function of $\phi$ in simulations (the red dots), which is compared with both rheological experiments (the blue dots) and data from previous papers \cite{srinivasan2021numerical,gamonpilas2016shear,gallier2014rheology,mari2014shear,boyer2011unifying,sierou2002rheology,zarraga2000characterization,lewis1968viscosity}, which is shown in Fig \ref{fig: Simulation results_eta1mu05d008}(d). The black dash-dot line represents the trajectory of $\eta_r$ in previous papers, while the blue dashed line characterizes the trajectory of experimental results and the red dashed line corresponds to simulation results with a finite $\dot{\gamma}$.  When $\phi < 0.5$, the relative viscosity $\eta_r$ increases with the solid fraction $\phi$ and follows the universal trajectory of rheological experiments and previous papers. The difference appears at $\phi > 0.5$, both experimental and simulated $\eta_r$ do not diverge at $\phi_c$, while the growth of it slows down with the increasing $\phi$. In this region, suspensions reach the dense packing state, and the cluster size only slightly increases as the solid fraction continues to rise, leading to a small stress increment. Consequently, the increment of the relative viscosity also decreases as $\eta_r = \eta_s / \eta_f$, where $\eta_s = \tau / \dot{\gamma}$ is the apparent viscosity of suspensions. 

It is significant that the behavior of suspensions is not only dependent on $\phi$, but also greatly influenced by $\dot{\gamma}$, and $\eta_r$, $\mu_e$, and $I_v$ are sensitive to $\dot{\gamma}$ (also see in Fig(6) in the supplementary file), particularly at higher $\phi$. In these cases, the rheological properties of suspensions cannot be fully characterized by traditional viscosity rheology and the $\mu_e-I_v$ rheology alone. To address this, we introduced a rescaling approach by normalizing the sheared relative viscosity and the shear stress with the distance to the transition point $|\phi_d - \phi|$. The rescaling results demonstrate a good collapse of data and describe the $\phi$-dependence and the $\dot{\gamma}$-dependence of granular suspensions in a universal framework.
\begin{figure}[ht!]
\includegraphics[scale=0.2]{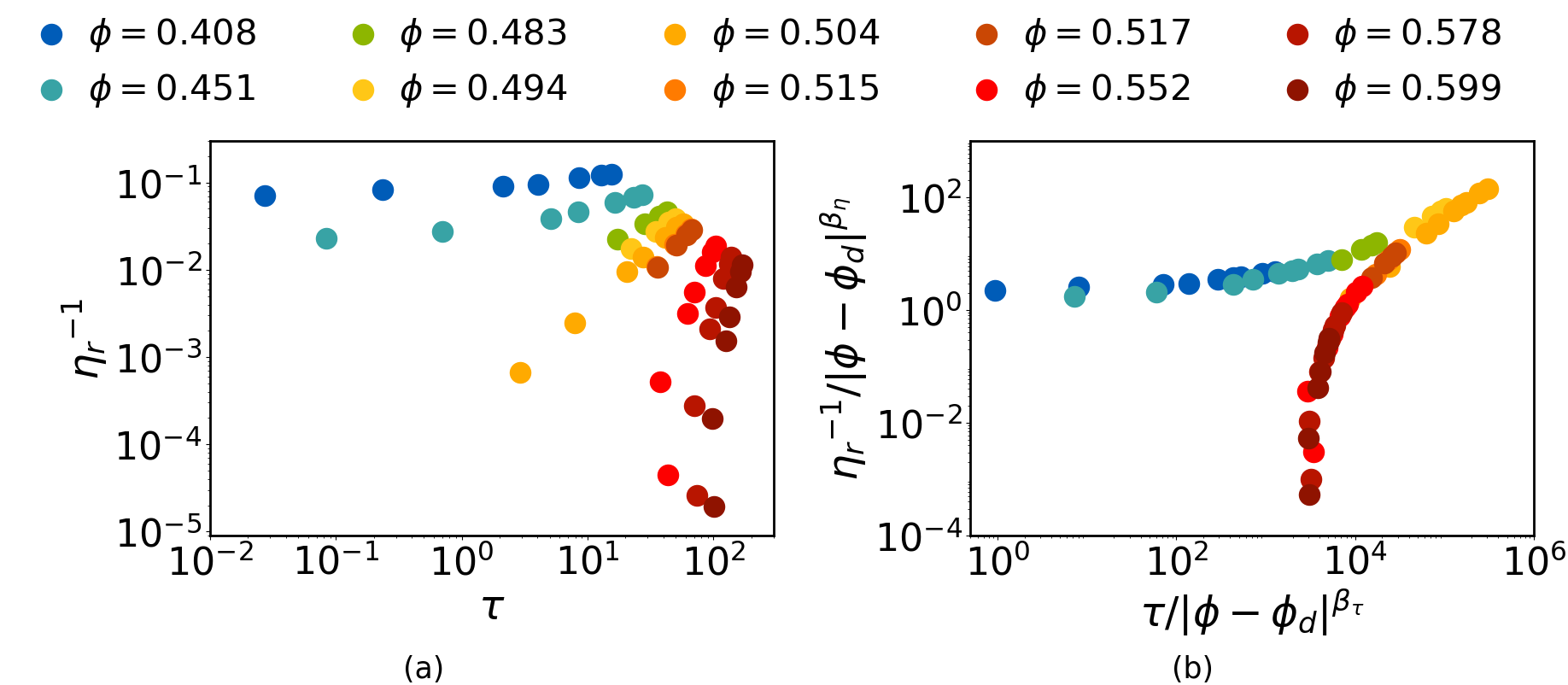}\\
\caption{\label{fig: Rescaling of relative viscosity} Critical scaling results of suspensions with $D_p = 0.071 \ \rm{mm}$, $\eta_f = 0.0206 \ \rm{Pa \cdot s}$, and $\mu_p = 0.5$. (a) The inverse relative viscosity ${\eta}_r^{-1}$ versus the shear stress $\tau$. (b) The rescaled curves of ${\eta}_r^{-1}$ and $\tau$ with ${|\phi_d - \phi|}^{\beta}$. $\beta_{\tau}$ and ${\beta_{\eta}}$ are the exponents of the yield stress and the inverse relative viscosity as the function of $|\phi_d - \phi|$.}
\end{figure}

Fig \ref{fig: Rescaling of relative viscosity}(a) shows the behavior of granular suspensions in relation to the inverse relative viscosity ${\eta}_r^{-1}$ and the shear stress in the log-log scale. Curves with different solid fractions have two distinct regions, concave curves for $\phi < 0.50$ and convex curves for $\phi > 0.50$. The transition point between these two regions corresponds to a transitional packing state with $\phi_d \approx 0.50$, which separates the region with $\phi < \phi_d$ as the dilute regime, and the region with $\phi > \phi_d$ as the dense regime. Within these two regimes, the development of ${\eta}_r^{-1}$ and $\tau$ are related to $|\phi - \phi_d|$. In the dilute regime ($\phi < \phi_d$), the increment of ${\eta}_r^{-1}$ is smaller than that of $\tau$ at a finite $\phi$. This indicates that $\tau$ is more sensitive to the increasing $\dot{\gamma}$ compared of $\eta_r$. Conversely, in the dense regime ($\phi > \phi_d$), ${\eta}_r^{-1}$ is more sensitive, and its growth is larger than that of $\tau$. Based on that, we investigate the relationship between $|\phi - \phi_d|$ and the inverse relative viscosity ${\eta}_{r}^{-1}$ in the dilute regime and the yield stress $\tau_y$ in the dense regime, given by
\begin{subequations}
\setlength{\abovedisplayskip}{3.2pt}
\begin{align}
    {\eta}_{r}^{-1} \sim {(\phi_d - \phi)}^{\beta_{\eta}}, \ \phi < \phi_d,\\
\tau_y \sim {(\phi -\phi_d)}^{\beta_{\tau}}, \ \phi > \phi_d,
\end{align}
\end{subequations}
where $\tau_y$ is the value at $\dot{\gamma} \to 0$ (See Eq(24) in the supplementary file). It was found that the values for these constant exponents are $\beta_{\eta} = 1.434$, and $\beta_{\tau} = 1.476$ (see Fig(7) in the supplementary file).

Based on these, we rescale ${\eta}_r^{-1}$ and $\tau$ with ${|\phi_d - \phi|}^{\beta_{\eta}}$ and ${|\phi_d - \phi|}^{\beta_{\tau}}$ separately, and observe the good collapse of Fig \ref{fig: Rescaling of relative viscosity}(b). In the dilute regime ($\phi< \phi_d$), the rescaled inverse relative viscosity ${\eta}_r^{-1}/{|\phi_d - \phi|}^{\beta_{\eta}}$ is a function of the rescaled $\tau/{|\phi_d - \phi|}^{\beta_{\tau}}$ with a single concave curve. At small $\phi$, suspensions behave as Newtonian fluids with small $\tau$, and the relative viscosity is independent of $\dot{\gamma}$ and is solely influenced by $\phi$. After accounting for the effect of $\phi$, the rescaled inverse relative viscosity tends to approach a constant value, where the viscosity is nearly equal to the fluid viscosity $\eta_f$ of the pure solvent as $\tau \to 0$ and $\phi \to 0$. With increasing $\phi$, suspensions show a $\dot{\gamma}$-dependence and tend to be non-Newtonian at high shear rates, ${\eta}_r^{-1}/{|\phi_d - \phi|}^{\beta_{\eta}}$ increases as the relative viscosity decreases with $\dot{\gamma}$, and becomes close to the maximum value at $\phi \to \phi_d$. In this regime, suspensions are primarily affected by $\phi$, with a significant increment of $\tau$, and gradually transition to non-Newtonian fluids with the influence of $\dot{\gamma}$. For $\phi > \phi_d$, suspensions reach the dense packing state, the increment of $\tau$ is caused by the increment of particle contact pairs decreases, while the influence of $\dot{\gamma}$ becomes more pronounced, resulting in strong shear thinning. ${\eta}_r^{-1}/{|\phi_d - \phi|}^{\beta_{\eta}}$ sharply increases within a small range of $\tau/{|\phi_d - \phi|}^{\beta_{\tau}}$ with a convex curve, and suspensions tends to have the maximum yield stress when the $\eta_r \to \infty$. After scaling, the relationship between the inverse relative viscosity and the shear stress can be described as
\begin{equation}
    {\eta}_r^{-1}=\left\{
    \begin{aligned}
    {{(\phi_d-\phi)}^{\beta_{\eta}}}f(\tau/{(\phi_d-\phi)}^{\beta_{\tau}}), \ \phi<\phi_d, \\
    {{(\phi-\phi_d)}^{\beta_{\eta}}}g(\tau/{(\phi-\phi_d)}^{\beta_{\tau}}), \ \phi>\phi_d.
    \end{aligned}
    \right
    .
    \label{eq: Rescaled equation}
\end{equation}

We apply the same method on suspensions with different fluid viscosities ($\eta_f$ = 0.00206, 0.0206, and 0.206 $\rm{P_a \cdot s}$ with $D_p = 0.071 \ \rm{mm}$ and $\mu_p = 0.5$, and $\eta_f$ = 0.0206 and 2.06 $\rm{P_a \cdot s}$ with $D_p = 0.893 \ \rm{mm}$ and $\mu_p = 0.5$) and different particle frictional coefficients ($\mu_p$ = 0.25, 0.5, and 1.0 with  $D_p = 0.893 \ \rm{mm}$ and $\eta_f$ = 0.00206 $\rm{P_a \cdot s}$). The results are shown in Fig \ref{fig: Rescaling of relative viscosity_All}(a) (also see Fig(8) in the supplementary file). When the particle size remains the same, suspensions with different $\eta_f$ and different $\mu_p$ exhibit similar trajectories and follow Eq \ref{eq: Rescaled equation} after being rescaled by ${|\phi-\phi_d|}^{\beta}$. However, the value of $\phi_d$ decreases with increasing $\mu_p$, which is 0.445 when $\mu_p = 1.0$ and 0.554 when $\mu_p = 0.25$ (insert figure in Fig \ref{fig: Rescaling of relative viscosity_All}(b)). The cluster size ratio $C_{max}$ at the transition point is also small for suspensions with large $\mu_p$ (see Fig 9(b) in the supplementary file). Large $\mu_p$ enhances the frictional contact between particles, forming a more stable structure with a stronger contact network. Consequently, suspensions acquire higher viscosity and shear stress at small $\phi$ with small particle clusters. These demonstrate that the dense packing state not only depends on the particle concentration or the cluster size, but also provides an assessment of the stability and strength of the contact structure within the suspension.
\begin{figure}[ht!]
\includegraphics[scale=0.2]{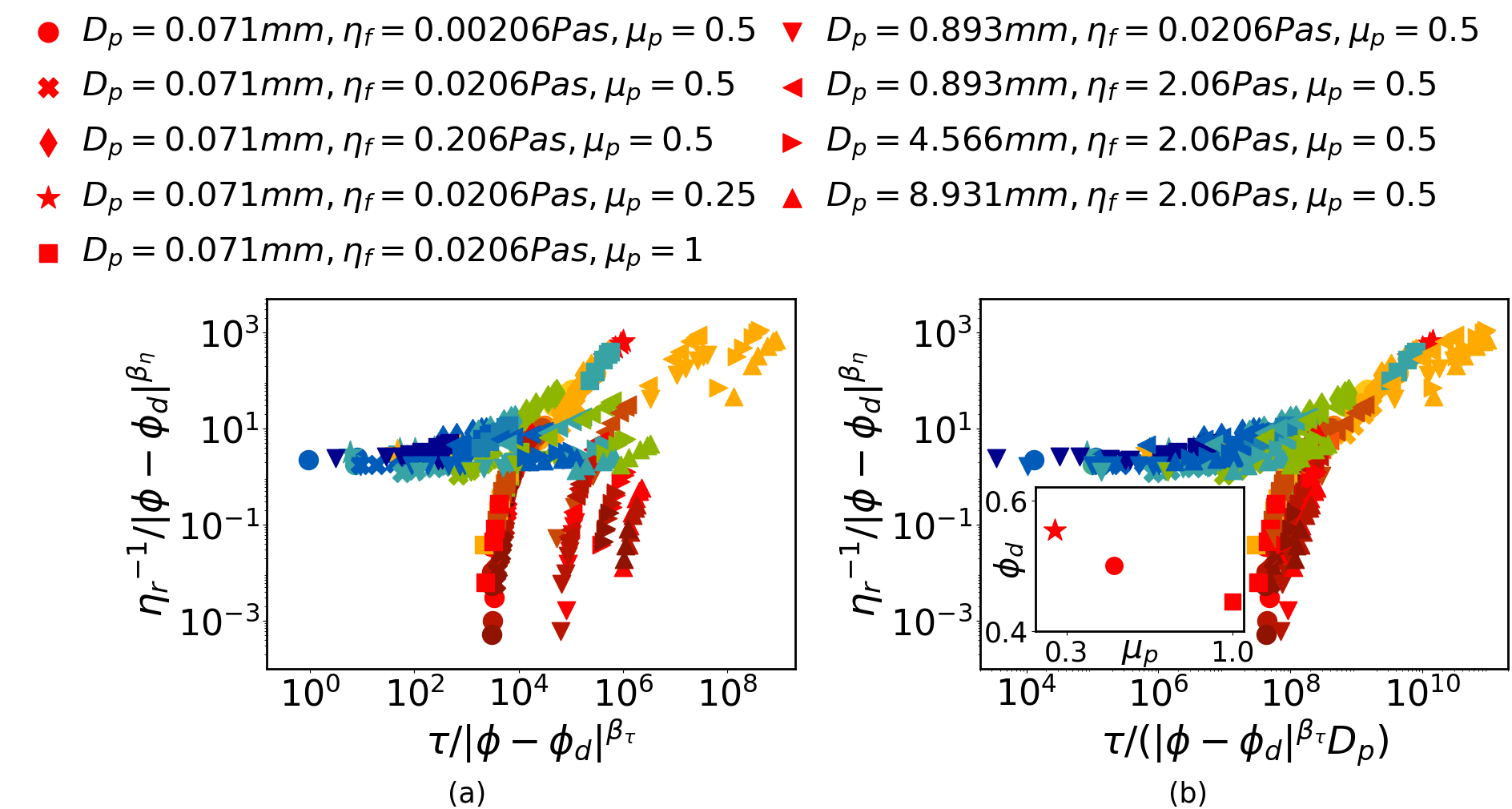}
\caption{\label{fig: Rescaling of relative viscosity_All}Rescaled inverse relative viscosity versus rescaled shear stress of all groups of suspensions, with different $D_p$, $\eta_f$, and $\mu_p$. (a) The rescaled curves of ${\eta}_r^{-1}$ and $\tau$ with $|\phi_d-\phi|^{\beta}$. (b) The rescaled curves considering of particle size effect, where ${\eta}_r^{-1}/|\phi-\phi_d|^{\beta_{\tau}}$ is a function of $\tau/(|\phi-\phi_d|^{\beta_{\tau}}D_p)$.}
\end{figure}
When we change the particle size of suspensions, the results show the difference in Fig \ref{fig: Rescaling of relative viscosity_All}(a). The ratio between the simulation space and the particle size is almost the same for suspensions with different particle sizes ($D_p$ = 0.071 mm, 0.893 mm, 4.566 mm, 8.931 mm with $\eta_f$ = 2.06 $\rm{P_a \cdot s}$ and $\mu_p = 0.5$). Under the same conditions, suspensions with larger particles have larger shear stress $\tau$, positioning their curves move toward the right region in the figure. To account for the effect of particle sizes, we rescale the shear stress as $\tau/({|\phi_d - \phi|}^{\beta_{\tau}}D_p)$. Subsequently, all curves move towards a universal collapse, with slight fluctuations, as depicted in Fig \ref{fig: Rescaling of relative viscosity_All}(b), and the equation is described by Eq \ref{eq: Rescaled equation with particle size},
\begin{equation}
    {\eta}_r^{-1}=\left\{
    \begin{aligned}
    {{(\phi_d-\phi)}^{\beta_{\eta}}}f(\tau/({(\phi_d-\phi)}^{\beta_{\tau}}D_p)), \ \phi<\phi_d \\
    {{(\phi-\phi_d)}^{\beta_{\eta}}}g(\tau/({(\phi-\phi_d)}^{\beta_{\tau}}D_p)), \ \phi>\phi_d.
    \end{aligned}
    \right
    .
    \label{eq: Rescaled equation with particle size}
\end{equation}
In this case, the effect of the solid fraction $\phi$, the shear rate $\dot{\gamma}$, and the particle size $D_p$ is described in a universal frame, suspensions with various parameters have similar transition behavior from the dilute to the dense regime.

We count the value of $\phi_d$ and the cluster size ratio $C_{max}$ at $\phi_d$ for all groups of suspensions in our tests (see Fig 9 in the supplementary file). As mentioned above, the value of $\phi_d$ is mainly affected by the particle frictional coefficient $\mu_p$, but only slightly increases at small fluid viscosity ($\phi_d$ = 0.51 with $\eta_f$ = 0.00206 $\rm{Pa \cdot s}$) and large particle size ($\phi_d$ = 0.507 with $D_p$  = 8.931 mm). Low-viscosity fluid ($\eta_f = 0.00206 \ \rm{Pa \cdot s}$, which is close to water) impacts a small drag force on solid particles, inducing a weakening contact network and smaller cluster size. At the same time, the contact between large particles is unstable, forming a smaller cluster. In these cases, suspensions achieve the dense packing state for a relatively large $\phi_d$, and more particles join in the contact network, leading to enough stress and viscosity at the transition point. For suspensions with different physical parameters, $C_{max}$ at $\phi_d$ is affected by both $\eta_f$, $\mu_p$, and $D_p$ as $C_{max} \sim \eta_f{\mu}_p^{-1}{D}_p^{-1}$.

In conclusion, across dilute to dense regimes, non-colloidal and non-Brownian granular suspensions with rigid spheres transform from Newtonian-like to non-Newtonian at the transition point $\phi_d$. Continuous shear thinning is observed in the DEM-LBM simulations, while suspensions show both strong $\phi$-dependence and $\dot{\gamma}$-dependence at the same time, especially at high solid fractions, which is associated with the development of the particle cluster. These two regimes are separated by the transition point $\phi_d$, and the behavior of suspensions is strongly related to the distance to $\phi_d$. We rescaled the inverse relative viscosity ${\eta}_r^{-1}$ and the shear stress $\tau$ with the distance from the transition point $|\phi-\phi_d|$ and found a universal collapse, with ${\eta}_f^{-1}/{|\phi-\phi_d|}^{\beta_{\eta}} \sim \tau/{|\phi-\phi_d|}^{\beta_{\tau}}$. With the consideration of particle size, the rheology of granular suspensions is characterized as ${\eta}_r^{-1}/{|\phi-\phi_d|}^{\beta_{\eta}} \sim \tau/({|\phi-\phi_d|}^{\beta_{\tau}}D_p)$. It provides a way to connect the relationship between macroscopic behavior and microscopic mechanisms and describes the $\phi$-dependence and the $\dot{\gamma}$-dependence in a uniform frame. 

We acknowledge the financial support from the National Natural Science Foundation of China with project numbers 12202367 and 12172305. We thank Westlake University, the Westlake High-performance Computing Center, and the Instrumentation and Service Center for Physical Sciences for computational and experimental sources and corresponding assistance.

\bibliography{citation}
\end{document}